# MODELING ORBITAL MODULATION OF CYG X-3 BY PARTICLE SIMULATIONS

**Osmi Vilhu[(1)], Pasi Hakala[(2)], Linnea Hjalmarsdotter[(1,3)], Diana Hannikainen[(1)], Ada Paizis[(4)], Michael McCollough[(5)]**

[(1)]Observatory,Box 14,FIN-00014 University of Helsinki, Finland, E-mail:osmi.vilhu@helsinki.fi
[(2)]Tuorla Observatory, University of Turku, Finland, E-mail:pahakala@astro.Helsinki.fi
[(3)]Observatory,University of Stockholm,Sweden, E-mail:nea@astro.helsinki.fi
[(4)]INAF-IASF, Milano, Italy, E-mail:ada@iasf-milano.inaf.it
[(5)]Smithsonian Astrophysics Observatory, USA, E-mail:mmccollough@head.cfa.harvard.edu

## ABSTRACT

The formation of the circumbinary envelope of Cygnus X-3 was studied by particle simulations of the WR (Wolf Rayet) -companion wind. Light curves resulting from electron scattering absorption in this envelope were computed and compared with observed IBIS/ISGRI and BATSE light curves. The matching was relatively good. For reasonable values of binary parameters (masses, inclination) and wind velocities, a stable envelope was formed during a few binary orbits. Assuming approximately $10^{-6}$ solar mass/year for the rate of the WR-wind, the observed light curves and accretion luminosity can be re-produced (assuming Thomson scattering opacity in the ionized He-rich envelope). The illuminated envelope can also model the observed *Chandra*-spectrum using the photoionizing XSTAR-code. Furthermore, we discuss observed radial velocity curves of IR emission lines in the context of simulated velocity fields and find good agreement.

## 1. INTRODUCTION

Cygnus X-3 is a close binary star with an orbital period of 4.8 hours. The donor star is probably a Wolf Rayet-star of WNE-type ([2,5]) whose wind feeds the accretion luminosity $10^{38-39}$ erg/s from radio to X-rays ([3]). The compact object can be either a neutron star or a black hole. Cyg X-3 displays a 4.8 hour modulation in X-rays and IR which can be assumed to be due to orbital motion (see e.g. [6,7]).

In this paper we study the possibility that this modulation is due to electron scattering in an asymmetric co-rotating circumbinary envelope. Pringle ([9]) was among the first to propose electron scattering to produce the observed smooth orbital modulation. We study the formation of this envelope by particle simulations from a spherical WR-wind, construct model light curves and compare them with the observed hard X-ray IBIS/ISGRI and BATSE light curves. In addition, the nebula radiates via excitation by the X-ray source and WR He-star. We construct a model spectrum using the XSTAR-code and compare it with a *Chandra* spectrum. We also discuss simulated velocity fields and observed IR emission line radial velocity curves.

## 2. PARTICLE SIMULATIONS

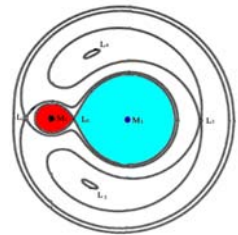

**Fig.1.** Binary potential surfaces

Non-interacting particle simulations of a spherical WR-wind with initial velocities V = 1200 – 1600 km/s were performed in the binary potential

$$\Psi = GM_1/r_1 + GM_2/r_2 + 0.5\Omega^2 r^2$$

(see Fig.1). The masses of the components were 10 solar masses for the donor WR Helium star, and 2 and 5 solar masses for the compact star (mass ratios q = 5 and 2). The WR-star radius of 0.92 solar radii ([8]) is inside its Roche lobe. We used 16-s time steps and the number of particles within the escape shell, 30 times the components' separation, was kept constant at 3 million particles. A new wind particle was generated when a test particle moved outside the escape shell, returned back to the WR-star or passed closer than 0.15 solar radii from the compact star (and accreted).

## 3. SIMULATION RESULTS

In all cases a stable circumbinary co-rotating envelope was formed after a few binary orbits. For q=5 and V=1200 km/s (see Figs 2 and 3), the envelope mass was replaced once per orbit giving $5.5 \cdot 10^{-10}$ solar mass for its mass (assuming mass loss rate $10^{-6}$ solar mass/year for the WR-wind). The accretion rate onto the compact star was $10^{-8}$ solar mass/year giving an accretion luminosity of $0.6 \times 10^{38}$ erg/s for 10 % efficiency. This leads to an average electron density $N_e = 2 \times 10^{12}$ cm$^{-3}$ and column density $7 \times 10^{23}$ cm$^{-2}$ (assuming a hydrogen deficient totally ionized envelope). These numbers are very similar to those estimated from the partial eclipse depth (50 %) of the hard X-ray light curves, assuming Thomson scattering opacity. For a higher wind velocity 1600 km/s the accretion rate onto the compact object was two times smaller and all numbers naturally scale with the assumed wind mass loss rate.

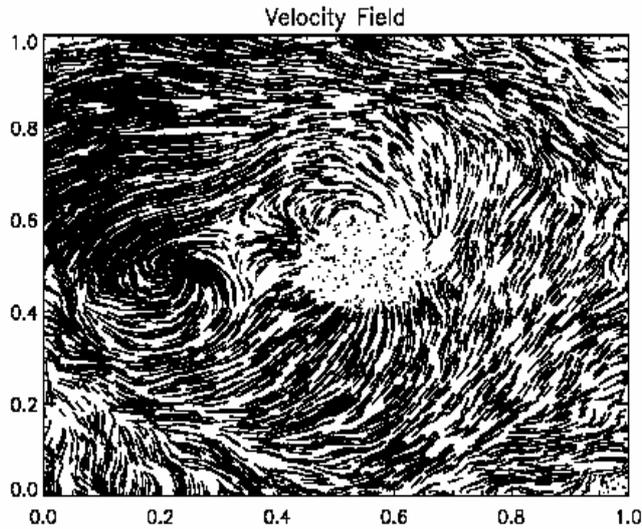

Fig.2. The velocity field in the orbital plane for q = 5 and V = 1200 km/s. The donor WR-star (10 solar mass) is seen in the center and the compact star (2 solar mass) to the left.

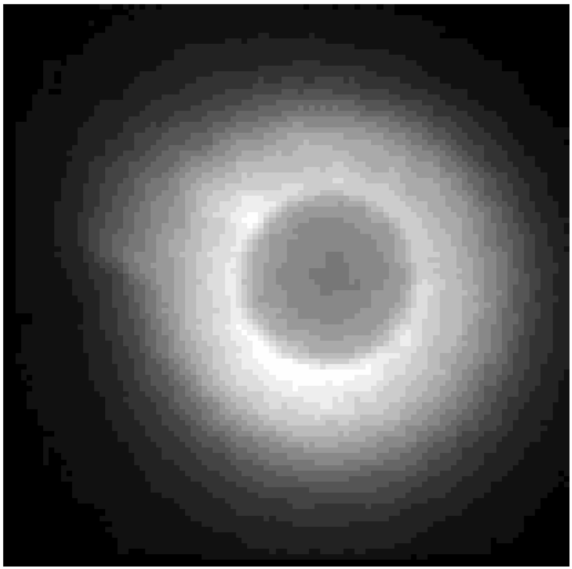

Fig.3. The projected particle density in the orbital plane for q = 5 and V = 1200 km/s. The donor WR-star (10 solar mass) is seen in the center and the compact star (2 solar mass) is to the left.

### 3.1. Comparison with hard X-ray light curves

Model light curves were computed for different viewing angles (inclinations) assuming that the optical depth depends on the line of sight column density exp(-const x column density) which is true for pure electron scattering opacity. These were then compared with the observed ISGRI 20-40 keV (revs 251-253, November 2-10, 2004) light curve, covering 10 contiguous orbits. Best agreement was achieved for high inclinations (Figs. 4 and 5). Due to the lack of a total eclipse the inclination can not be much larger than 60 degrees.

The small asymmetry of the observed IBIS/ISGRI light curve does not follow from our models, yet, resulting in relatively high reduced $\chi^2$ - values (see Fig.4). The soft X-ray ASM light curve (2 – 12 keV) is even more asymmetric (Fig.6), probably due to photoelectric absorption in partially ionized gas, in the rising part of the light curve.

The mean BATSE (20-80 keV) light curve is not so asymmetric (Figs. 6 and 7). It is possible that the small asymmetry in the IBIS/ISGRI light curve is not permanent. The similarity of the BATSE light curves during states with low (< 10 cps) and high (10-35 cps) levels of the ASM 2 – 12 keV flux (Fig.7) indicates that the circumbinary envelope does not depend on the X-ray luminosity. This is the case if the WR-wind feeds the envelope with small radiation pressure effects like in our modelling.

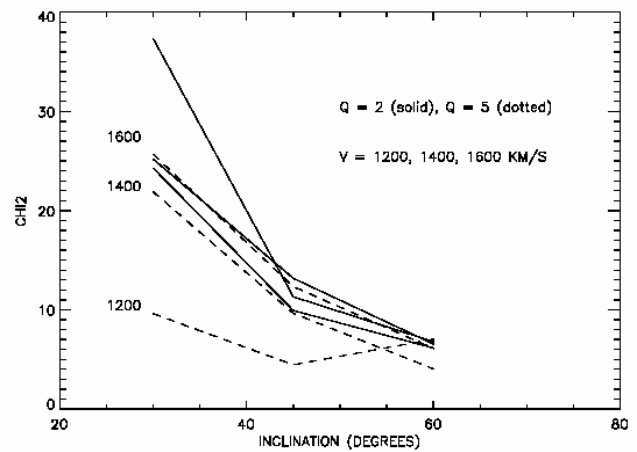

Fig. 4. Reduced $\chi^2$-values between the models and ISGRI light curves vs inclination, for different mass ratios and wind velocities.

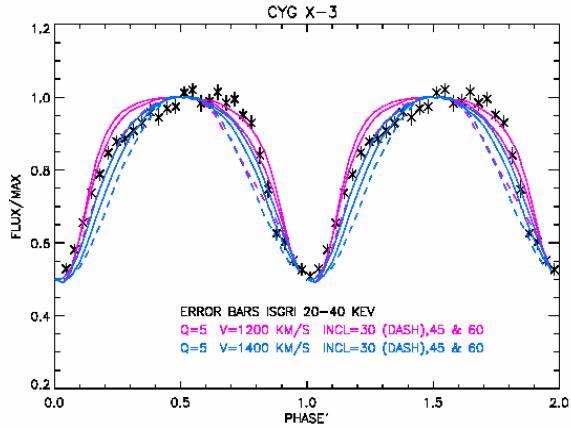

Fig.5. Model light curves compared with the IBIS/ISGRI 20-40 keV light curve (crosses with error bars).

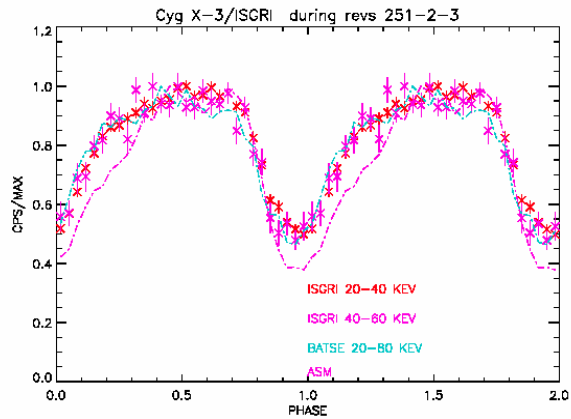

Fig. 6. IBIS/ISGRI (red and pink crosses), BATSE (blue dashed line) and ASM (pink dashed line) light curves compared.

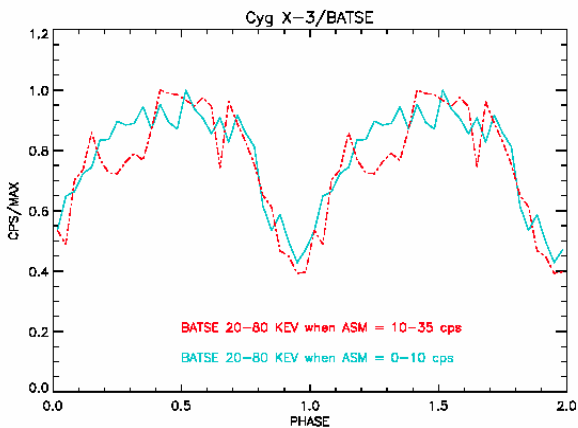

Fig.7. BATSE light curves during the ASM low (solid blue line) and high (red dashed line) states.

### 3.2. Comparison with CHANDRA spectrum

The simulations also predict a nice agreement with the *Chandra*-spectrum taken during a low state ([4]), retrieved from the archive (acis1456). (Fig.8). The predicted spectrum (transmitted and outward) was computed with the XSTAR-code using abundances of a weak line WN3-4 star as given in [1]: H=0.1, He=10, C=0.56, N=40, O=0.27 and other elements =1, relative to the solar ones. The electron density $n_e = 10^{12}$ cm$^{-3}$ and column density $N_H = 3.5 \times 10^{23}$ cm$^{-2}$ were used in the spherical He-rich envelope, to mimic the simulated circumbinary (non-spherical) envelope.

Two ionizing radiators were assumed: the X-ray source and the Helium (WR) star. The X-ray radiator spectrum was computed as a mean of two typical low/hard state RXTE/PCA spectra (xp30082040200 and xp30082040400) as fitted with the wabs*(compPS+Gauss) XSPEC-model, giving unabsorbed bolometric luminosity $0.83 \times 10^{38}$ erg/s. The model includes photoelectric absorbtion, weak disk BB, Comptonized and reflection components and the Iron line. The Helium-star radiator was assumed to be a black body with effective temperature of $10^5$ K and luminosity $5 \times 10^{38}$ erg/s.

As seen, the agreement is good, predicting the the Sulphur and Silicon K$\alpha$ lines as well as the continuum. The effect of the He-star was to produce extra heating and increase of the gas temperature and, hence, lower the overall line emissivity. Without the soft He-star radiator the lines in Fig.8 would be stronger.

We note that the model nebula may also explain the IR-spectrum and K-magnitude of Cyg X-3 if one adds the He-star to excite the nebula with some IR-excess (in prep.).

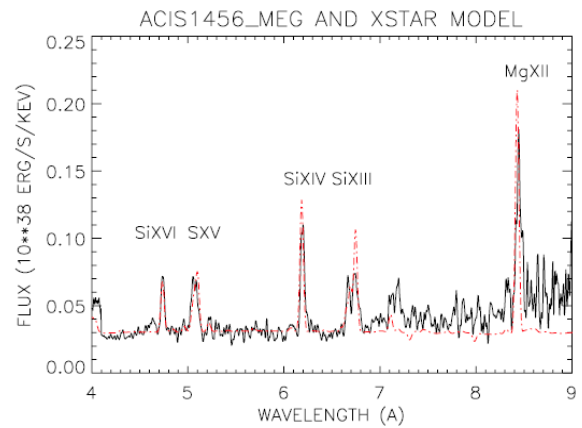

Fig.8. *Chandra* low state spectrum (black line, de-reddened by NH = $5 \times 10^{22}$ cm$^{-2}$) and corresponding model spectrum (red dash-dot line).

## 4. DISCUSSION AND CONCLUSIONS

It is interesting to see that simple non-interacting particle simulations can qualitatively reproduce the observed hard X-ray light curves for reasonable values of masses and wind velocities, although the results depend rather heavily on these. However, the small asymmetry of the IBIS/ISGRI light curve during INTEGRAL revolutions 251 - 253 remains unexplained (Fig.5). Best fitting light curves are produced in our numerical experiments for a compact star mass of 2 solar mass (assuming 10 solar mass for the WR-star), wind velocities 1200 – 1400 km/s and large inclinations.

Using the photoionization code XSTAR it was possible to reproduce the low state *Chandra* spectrum without any extra adjustments of parameters.

The concept presented here predicts light curves which are independent on the X-ray source luminosity and depend just on the WR-wind. In fact this seems to be the case (Fig. 7). For observed WR-star mass loss rates $10^{-5}$ – $10^{-6}$ solar mass/year, the simulations predict accretion rates of correct order of magnitude to explain the luminosity $10^{38}$ erg/s.

In principle, radial velocities could be used to test our models as these predict velocity fields including the binary motion (see Fig. 2). The problem is to identify lines and regions where they are formed. This would require knowledge about line formation in the envelope which we have not attempted. Ref. [2] derived a mass function of 0.027 solar mass assuming a Helium infrared (IR) absorption feature to arise in the WR-star wind. The amplitudes of the IR HeII and NV emission lines were much larger (up to 400-500 km/s) and 1/4 of an orbit out of phase with the absorption lines, having maximum blueshifts around the X-ray phase 0.

It is possible that these IR emission lines are formed in the shadow region behind the WR-star. Elsewhere Helium and Nitrogen are probably fully ionized due to the strong X-ray source. The radial velocity curve was computed from the WR-shadow of the model in Fig. 2. The result is shown in Fig. 9 for separate shells with distances of 1–30 solar radii from the WR-star. The average formation region of the IR HeII 2.2 micron line is around the distance of 2 – 4 solar radii from the WR-star (as estimated with the XSTAR code using hot He-star radiator). This mean velocity curve is shown in Fig.9 with red crosses. The amplitude is roughly what observed, and the maximum blueshift is around phase 0 as observed.

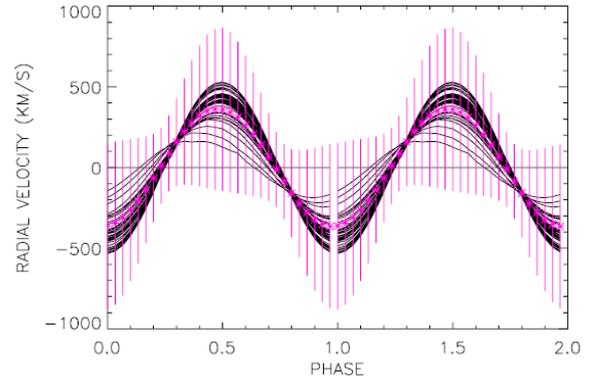

Fig.9. Radial velocity curves of IR- emission lines from separate shells in the X-ray shadow region behind the WR-star, according to the model shown in Fig. 2. The red curve (crosses) show the mean velocity curve weighted with the HeII 2.2 micron line emissivity and gas density. The red error bars show standard deviation of particle velocities from the mean value.